
%
%
\magnification=\magstep1
\baselineskip=16pt
\overfullrule=0pt

\def\G{{\cal G}}
\def\H{{\cal H}}
\def\M{{\cal M}}
\def\R{{\bf R}}
\def\half{{1 \over 2}}

\def\tr{{\rm tr}}
\rightline{UCLA/95/TEP/11}
\vglue0.9cm
\centerline{{\bf INVARIANT EFFECTIVE ACTIONS AND COHOMOLOGY}
\footnote{*}{ Research supported in part by NSF grant PHY-92-18990.}}
\vglue1.5cm
\centerline{{\bf Eric D'Hoker}}
\medskip
\centerline{{\it Physics Department}}
\centerline{{\it University of California, Los Angeles}}
\centerline{{\it Los Angeles, CA 90024, USA}}
\medskip
\centerline{\rm E-mail: dhoker@physics.ucla.edu}

\vglue1.5cm
\centerline{\bf Abstract}
\vglue0.7cm

We review the correspondence between effective actions resulting
from non-invariant Lagrangian densities, for Goldstone bosons arising
from spontaneous breakdown of a symmetry group G to a subgroup H,
and non-trivial generators of the de Rham cohomology of G/H.
We summarize the construction of cohomology generators in terms of
symmetric tensors with certain invariance and vanishing properties
with respect to G and H. The resulting actions in four dimensions
arise either from products of generators of lower degree such as the
Goldstone-Wilczek current, or are of the Wess-Zumino-Witten type.
Actions in three dimensions arise as Chern-Simons terms evaluated on
composite gauge fields and may induce fractional spin on solitons.

\vglue1cm

{\it Contribution to the Proceedings of STRINGS 95, held at University
of Southern California, March 13 - 18, 1995.}

\vfill\eject

\leftline{\bf 1. Introduction}

\bigskip

We review recent investigations ${}^{1,2,3}$ into the structure of
effective actions for Goldstone bosons that arise in the process of
spontaneous breakdown of a continuous internal symmetry group $G$ to a
subgroup $H$. The universal nature of spontaneous symmetry breakdown
strongly constrains the form of the effective action of the
Goldstone fields. The low energy dynamics is completely determined by
the groups $G$ and $H$, and by only a finite number of coupling
constants, up to any given order in an expansion in powers of
derivatives or momenta.

Long ago, a general method was developed ${}^4$ for constructing
the most general invariant Lagrangian densities for the Goldstone
fields. Invariance of the action does not however, in general,
require invariance of the Lagrangian density, which may change
instead by a total derivative term. The Wess-Zumino-Witten (WZW)
term, which was originally considered as an effective action for
chiral anomalies, is an example of such an exception ${}^{5,6}$.

The success of the effective field theory approach depends critically
on the ability to enumerate all invariant terms in the effective
action with a given number of derivatives. Omission of certain terms
would lead to an inconsistent correspondence with the dynamics of the
underlying microscopic theory and would invalidate the effective
action approach. Thus, we are led to investigate the structure of the
most general actions, including those that do not arise from
$G$-invariant Lagrangian densities. We shall leave the dimension of
space-time arbitrary.

In Ref. 1, we characterized effective Lagrangian densities that,
although not $G$-invariant, yield $G$-invariant effective actions,
in terms of the de Rham cohomology of the coset space
$G/H$.  The corresponding invariant effective actions $S[\pi]$ for the
Goldstone fields $\pi(x)$ in space-time dimension $n-1$ are
given in terms of the cohomology generators $\Omega $ of
degree $n$  by
$$
S[\pi] = \int _{B_n} \Omega (\tilde \pi)
\eqno (1.1)
$$
Here, $n-1$ dimensional space-time $M_{n-1}$ is extended to a ball
$B_n$ with boundary $\partial B_n =M_{n-1}$, and the field $\tilde
\pi$ interpolates continuously between the original field $\pi$ on
$M_{n-1}$ and the 0 field. Each non-trivial cohomology generator
produces a $G$-invariant action given by (1.1) which arises  from a
non-invariant Lagrangian density and escapes the construction of
$G$-invariant actions given in Ref 4.  The derivation of this
correspondence will be briefly reviewed in \S 2.

The de Rham cohomology is well-known in the case where $G/H$ is itself
a Lie group ${}^7$ and the results will be listed in \S 3.1.
For general coset spaces $G/H$, the structure of de Rham cohomology
has been the subject of intense study in mathematics ${}^8$. Yet, the
number and the form of the generators does not seem to be available
explicitly for general $G/H$. In Ref. [2], we obtained a
simple construction of all de Rham cohomology generators for
arbitary compact $G$ and subgroup $H$ in terms of symmetric tensors
with certain invariance and vanishing properties under $G$ and $H$.
This construction will be summarized in \S 3.2 and \S 3.3.

The invariant actions produced by this construction have direct
physical interpretations. For cohomology of degree 2, they describe
the dynamics of charged particles in a magnetic monopole field
${}^6$; for degree 3 they correspond to the WZW term in 2
dimensions. For degree 4, they may be recast in terms of the
Chern-Simons invariant ${}^{9,10}$ evaluated on an $\H$-valued gauge
field built out of the Goldstone fields and they provide
generalizations of the Hopf invariant ${}^{11}$ to arbitrary $G/H$.
For degree 5, they are WZW terms on cosets $G/H$, which
arise only when there exists a non-vanishing $G$-invariant symmetric
tensor of rank 3 (the $d$-symbols of chiral anomalies ${}^{10}$)
which vanishes upon restriction to the Lie subalgebra of $H$. The
fact that the WZW term can be constructed this way has been known for
some time${}^{12,13}$. What was shown in Refs. 1 and 2 is that this
construction produces {\it all Lagrangian densities} that, although
not $G$-invariant, yield $G$-invariant actions. The arguments are
reviewed in \S 4.
Coupling constant quantization in arbitrary dimension and
fractionalization of soliton spin in 2+1 dimensions are
discussed briefly in \S 5.

\bigskip
\bigskip

\leftline{\bf 2. Characterizing Invariant Actions by Cohomology of
G/H}

\bigskip

Let $G$ be a compact connected Lie group and $H$ a subgroup, with
Lie algebras $\G$ and $\H$ respectively. $\G$ may be decomposed as
$\G = \H + \M$, with $[\H, \H ] \subset \H$, and $[\H,\M]\subset\M $.
The Goldstone fields
\footnote{*}{We shall denote indices for the generators of $\G$,
$\H$ and $\M$ by capital $A=1, \cdots ,\dim G$, lower case Greek
$\alpha$  and lower case  $a $ respectively.}
$\pi ^a(x)$ transform under linear representations of $H$, but under
non-linear realizations of $G$. They parametrize the coset $G/H$ in
terms of the field $U(\pi)$ in $G$ which, under a global
transformation $g\in G$, maps $\pi\rightarrow \pi'$, with
$$
g\,U(\pi)=U(\pi')\,h(\pi,g)
\eqno (2.1)
$$
where $h(\pi,g)$ is some element of the unbroken subgroup $H$.
The transformation properties of the Lie algebra valued derivatives
follow from Eq. (2.1)
$$
U^{-1}(\pi'){\partial U(\pi') \over \partial \pi^a}=
h(\pi,g)\,U^{-1}(\pi){\partial U(\pi) \over \partial
\pi^a}\,h^{-1}(\pi,g)-{\partial h(\pi,g) \over \partial
\pi^a}\,h^{-1}(\pi,g)
\eqno (2.2)
$$
The $\H$ component of Eq. (2.2) transforms as an $\H$-valued gauge
field, while the $\M$ component transforms homogeneously. This allows
us to introduce a composite gauge field $V=V_\mu dx^\mu$, and an $\H$
covariant derivative $D^\H=D^\H _\mu dx^\mu$ as follows
$$
V=(U^{-1}dU)_\H \qquad \qquad U^{-1}D^\H U=
(U^{-1}dU)_\M = U^{-1}(dU -UV)
\eqno (2.3)
$$
The most general local invariant Lagrangian density is obtained as a
sum of monomials each of which is an invariant product of covariant
derivatives [4]
$$
{\cal L} = {\cal L}(U^{-1}D^\H _\mu U, \cdots, U^{-1} D^\H _{\mu_1}
\cdots D^\H _{\mu _n}U, \cdots ~)
\eqno (2.4)
$$
The field strength $W=dV+V^2$ of the gauge field $V$ does not have
to enter Eq. (2.4), since it is expressible in terms of the
commutator of two $D^\H$'s.

We now consider an action $S[\pi]$ which is invariant under $G$, so
that $S[\pi'] = S[\pi]$, but which is not necessarily obtained
from an invariant Lagrangian density. Still, we show that its
variation with respect to $\pi$ is an invariant density.  First, the
variation under an arbitrary change in $\pi$ may always be
written as
$$
\delta S[\pi]=\int _{M_{n-1}}d^{\scriptstyle n-1} x\;\tr
\,\left\{(U^{-1}\delta U)_\M\,J\right\}
\eqno (2.5)
$$
where $J$ is a local ${\cal M}$-valued function of $\pi$ and its
derivatives.  Since $S[\pi]=S[\pi']$ for all $\pi$, the variational
derivatives with respect to $\pi$ are also equal for all $\pi$
$$
\tr \,\left\{\left[U^{-1}(\pi'){\partial
U(\pi') \over \partial \pi^a}\right]_\M \,J(\pi')\right\}=\tr
\,\left\{\left[U^{- 1}(\pi) {\partial U(\pi) \over \partial
\pi^a}\right]_\M \,J(\pi)\right\}
\eqno (2.6)
$$
In view of Eq. (2.2), the $\M$ component of $U^{-1}\partial
U/\partial \pi ^a $ transforms homogeneously under $G$ by conjugation
with $h(\pi,g)$. Thus $J$ must also transform homogeneously in
order to satisfy Eq. (2.6). The $\M$ component of
$U^{-1}\delta U$ in Eq. (2.5) transforms homogeneously, again in
view of Eq. (2.2), so that the variation of the Lagrangian density
$\tr\{(U^{-1}\delta U )_\M J\}$ is invariant, as promised.
It follows that the contribution from any term in the invariant
effective action to the classical equations of motion will be
manifestly covariant under $G$.

 The above result leads to a natural $n$-dimensional formulation
of the invariant action.
As described in \S 1, $n-1$ dimensional space-time $M_{n-1}$ is
extended to an $n$ dimensional manifold $B_n$, of which $M_{n-1}$ is
the boundary. We introduce a smooth function $\tilde{\pi}^a(x,t^1)$,
which interpolates between $\tilde{\pi}^a(x,1)=\pi^a(x)$, and
$\tilde{\pi}^a(x,0)=0$. (Often at this point, space-time is
compactified to a sphere $S^{n-1}$, so that $B_n$ is an $n$-ball.
Here however, we shall not specify any particular topology for
$M_{n-1}$, as there may be physically interesting situations for
toplogies other than spherical. Our arguments apply in the case of
general topology as long as the interpolation $\tilde \pi$ exists
${}^1$.) The action may then be written in
$n$ dimensions
$$
S[\pi]=\int_{B_n}dx^{\scriptstyle n-1}\, dt^1\; {\cal L}_1\
\eqno (2.7)
$$
where ${\cal L}_1$ is the $G$-invariant density
$\tr \left\{(U^{-1}\,\partial U/\partial t_1)_\M J\right\}$.

The most general expression of this type can be obtained as in Eq.
(2.4), except for the fact that the density ${\cal L}_1$ must
satisfy the integrability conditions that guarantee that it arises
from a variation of an action.
Consider a general deformation $\pi(x)\rightarrow
\tilde{\pi}(x;t^i)$, where $t^i$ are a set of dim$~G/H -(n-1)$ free
parameters, that along with the $x^\mu$ provide a set of coordinates
for $G/H$.  We have shown that
$$
{\partial S[\tilde{\pi}] \over \partial t_i}=\int_{M_{n-1}}
d^{\scriptstyle n-1} x\;{\cal L}_i
\eqno (2.8)
$$
where ${\cal L}_i$ are $G$-invariant functions of $\tilde{\pi}^a$ and
its derivatives. Integrability of this system requires that
$$
{\partial{\cal L}_i \over \partial t^j}
-{\partial{\cal L}_j \over \partial t^i}=-\partial_\mu
{\cal L}^\mu_{ij}
\eqno (2.9)
$$
These equations have further integrability conditions on ${\cal
L}^\mu _{ij}$. It was shown in Ref. 1 that the entire sequence of
integrability conditions may be expressed as the closure of a
differential form $\Omega (\tilde \pi)$ of degree $n$ with respect to
$d = dt^i \partial _i + dx^\mu \partial _\mu $
$$
d\Omega (\tilde \pi)   =0 \qquad \qquad \qquad
\Omega (\tilde \pi)  = {\cal L}_i dt^i d^{\scriptstyle n-1}x +
\half {\cal L}_{ij} ^\mu dt^i dt^j d^{\scriptstyle n-2}x_\mu
\cdots
\eqno (2.10)
$$
Choosing one particular coordinate $t^1$ in the definition of the
integral (2.7), while keeping all the others fixed, the action may
be written as in Eq. (1.1).

When $\Omega $ is exact, $\Omega = dQ$, with $Q=Q_0
d^{\scriptstyle n-1}x + Q_i ^\mu dt^i d^{\scriptstyle n-2}x_\mu +
\cdots $ and $Q_0$ invariant under $G$, we recover the invariant
actions obtained from invariant Lagrangians, as given by Eq. (2.4).
Thus, all invariant actions {\it that are not associated with
invariant Lagrangian densities} are given by closed forms $\Omega$
modulo exact forms $dQ$, in which the leading terms ${\cal L}_i$ and
$Q_0$ respectively are $G$-invariant.

Two closed differential forms that are continuously connected to each
other differ by an exact form ${}^{14}$.  By construction, the form
$\Omega$ is invariant under reparametrizations of $t^i$, while under
infinitesimal reparametrizations of $x$ it changes by an exact
form of the type $dQ$, which we mod out by.  Thus, the class of forms
$\Omega$ up to exact forms is reduced to that of reparametrization
invariant $\Omega$'s, which are now well-defined  differential
forms on $G/H$, to be taken modulo forms $dQ$ where $Q$ is
also a well-defined form on $G/H$.  Furthermore, since
$G$ acts transitively on $G/H$, the $G$ transform of $\Omega$ is
continuously connected to the original form as well.  Since $G$ is
compact, one can construct a $G$-invariant form by integrating  over
$G$ with the invariant Haar measure ${}^{14}$. This has no effect on
(2.7) since the leading term ${\cal L}_i$ was already
$G$-invariant.  Also, one  similarly shows that any two
continuously connected invariant forms on $G/H$ differ not only by an
exterior derivative, but by the exterior derivative of
a {\it G-invariant} form $Q$. The space of closed $G$-invariant
well-defined forms on $G/H$ modulo the exterior derivative of
$G$-invariant well-defined forms on $G/H$ can be identified
${}^{14}$ with the de Rham cohomology of $G/H$ of degree $n$. Thus,
the classification of $G$-invariant terms in
$S[\pi]$ that do not correspond to an invariant Lagrangian
density is reduced to finding the $n$-th de Rham cohomology group
$H^n(G/H;{\bf R})$ of the homogeneous space $G/H$. Similar issues
were addressed independently from the point of view of
equivariant cohomology in Ref. 15.

\bigskip
\bigskip
\leftline{\bf 3. Cohomology of Homogeneous Spaces }

\bigskip

There are two ways in which the study of cohomology can be reduced.
First, for a product of spaces $K_1 \times K_2$, we
have the K\"unneth formula ${}^{14}$
$$
H^n (K_1 \times K_2;{\bf R})
= \sum _{n_1+n_2=n} ^{} H^{n_1}(K_1;\R) \wedge H^{n_2}
(K_2;{\bf R})
\eqno (3.1)
$$
which gives $H^n(G/H;\R)$ in terms of the cohomology of degree 0 up
to $n$ of its factors. Second, when a generator of degree $n$ can be
written as a linear combination of products of cohomology generators
of degree strictly less than $n$, it is called {\it decomposable}. Any
generator which does not contain any decomposable components is said
to be {\it primitive}. Thus, the entire cohomology can be
reconstructed from the primitive generators on the factor spaces of
$G/H$.

\vglue1cm
\leftline{\it 3.1. Cohomology for the case of G/H a Lie group}
\vglue0.4cm

When $G/H$ is itself a compact Lie group, it factors into simple
Lie groups and a number of ~$U(1)$ factors. Thus the cohomology
is obtained from the primitive generators on $U(1)$ and on all simple
Lie groups. We denote a primitive generator of degree $k$
by $\Omega_k$ with $\Omega_0=1$.
The list of primitive generators for the classical Lie groups is given
by the following catalog$^7$, where $ H^*(G;\R)$ denotes the
cohomology ring of $G$. We have $H^* (U(1);\R)  =
\wedge\{1,~\Omega_1\} $ and
$$
\eqalign{
H^* (SU(N);\R) =& \wedge \{1,~ \Omega_{2k+1}, ~k=1,~2, \cdots, N-1\}
\cr
H^* (SO(2N);\R) =& \wedge \{1,~ \Omega_{4k-1}, ~k=1,~2, \cdots, N-1;
	~\Omega'_{2N-1} \}  \cr
H^* (SO(2N+1);\R) =& \wedge \{1,~ \Omega_{4k-1}, ~k=1,~2,\cdots, N\}
\cr
H^* (Sp(2N);\R) =& \wedge \{1,~ \Omega_{4k-1}, ~k=1,~2,\cdots, N\}
\cr }
\eqno (3.2)
$$
The primitive generators are obtained from the left-invariant one
forms $U^{-1}dU$.
$$
\Omega _k = {1 \over  k!} \tr ~\{U^{-1}dU\} ^k
\eqno (3.3)
$$
Making use of the cyclicity property of the trace and the fact that
we have a power of a form of degree 1, we see that $\Omega _k=0$
whenever $k$ is an even integer. Antisymmetry of $U^{-1}dU$ for
$SO(N)$ groups (combined with symplectic conjugation for
$Sp(2N)$) furthermore implies that for these groups $\Omega _k=0$
whenever $k=1$ mod 4. The doubling of the generator of degree $4M-1$
for $SO(4M)$ is related to the existence of self-duality in those
dimensions. Thus, there are {\it two} non-trivial invariant actions in
dimension $d=4M-2$ for the group $SO(4M)$. The cohomologies of the
exceptional groups may also be found in Ref. 7; the degrees of the
primitive generators of $E_8$ for example are 3, 15, 23, 27, 35,
39, 47 and 59 ! In the case of 4 space-time dimensions,
$H^5(G;{\bf R})$ produces WZW  terms : if $G$ is semi-simple with
$p$ factors $SU(N_i)$ with $N_i\geq 3$ and all other factors with
$H^5=0$, we have $p$ different WZW terms, each with an independent
coupling constant.

\vglue1cm
\leftline{\it 3.2. Differential Calculus on Homogeneous Spaces
G/H}
\vglue0.4cm

It is a fundamental result ${}^{14}$ that the cohomology of
homogeneous spaces $G/H$  is given by the classes of closed
$G$-invariant forms on $G/H$ modulo forms that are the
exterior derivative of
$G$-invariant forms on $G/H$. The space of all $G$-invariant forms
in turn, is easily described in terms of left differentials $\theta
^A$, defined by
\footnote{*} {Here, $T^A$ are matrices in the representation of $U$,
$f$ are the structure constants of $G$ and $\theta$ is a flat
connection satisfying the Maurer Cartan equation $d\theta
+\theta ^2 =0$.}
$$
\theta = \theta ^A T^A = U^{-1}dU, \qquad\qquad [T^A,T^B]=f^{ABC}T^C
\eqno (3.4)
$$
A general form $\Omega$ of degree $n$
$$
\Omega = {1 \over n!} \omega _{A_1 \cdots A_n} \theta ^{A_1} \cdots
\theta ^{A_n}
\eqno (3.5)
$$
is $G$-invariant, provided the coefficients $\omega _{A_1 \cdots A_n}$
are constant, vanish whenever one of the indices $A_i$
corresponds to a generator in $\H$ and are invariant under the adjoint
action of the group $H$. This is easily understood in view of
the fact that the $\H$ components of $\theta$ transform as a gauge
field under $G$, and the tensor $\omega$ must vanish on $\H$ to
guarantee gauge invariance.
We also need differential forms $\Omega _{B_1\cdots B_m}$ of degree
$n$ which are tensors of rank $m$  on $\G$.

In addition, we define four standard operations ${}^{14}$ acting on
these forms, by
$$
\eqalign{
D\Omega _{B_1 \cdots B_m} = & ~d\Omega _{B_1 \cdots B_m} +
f_{B_1 BC} \theta _B \Omega _{C B_2 \cdots B_m} + \cdots +
f_{B_m BC} \theta _B \Omega _{ B_1 \cdots B_{m-1}C}
\cr
i_A \Omega _{B_1 \cdots B_m} = & ~1/ (n-1)!~ \omega _{B_1 \cdots
B_m;A A_2 \cdots A_n} \theta ^{A_2} \cdots \theta ^{A_n}
\cr
L_A \Omega _{B_1 \cdots B_m}  =& ~f_{AB_1B}\ \Omega _{B B_2
\cdots B_m}  	+ \cdots + f_{AB_mB}\ \Omega _{B_1 \cdots B_{m-1} B}
\cr
\Delta \Omega _{B_1 \cdots B_m} =& ~L_A (i_A \Omega _{B_1 \cdots B_m}
)
\cr}
\eqno (3.6)
$$
The (covariant) exterior derivative $D$ satisfies $D^2=0$ and
increases the degree by one unit while leaving the rank  unchanged.
The interior product $i_A$ satisfies $i_A i_B + i_B i_A=0$, and
increases the rank and lowers the degree by 1. The $\G$ rotation
$L_A$ acts as a derivative and its square
$L^2=L_AL_A$ is the quadratic Casimir operator. The operation
$\Delta$ lowers the degree by 1 while keeping the same rank; it is
the adjoint operator of $D$.
The operations $D$ and $\Delta$ commute with $L_A$, while $i_A$ and
$L_A$ transform in the adjoint representation of $\G$. Two
additional relations between these operations
$$
i_A D   + D i_A  =  - L_A
\qquad \qquad \qquad
D\Delta  + \Delta D  =  -L^2
\eqno (3.7)
$$
form the cornerstone of the analysis of the cohomology of
$G/H$.

\vglue1cm
\leftline{\it 3.3. Cohomology of General Compact Homogeneous Spaces
$G/H$ : an Outline}
\vglue0.4cm

It is a standard result that the cohomology of $G/H$ can be obtained
from the solution of a purely group theoretic problem ${}^8$. The
explicit construction of generators does not seem to be available
however, and will be outlined here. A detailed account is given in
Ref. 2. As mentioned before, all cohomology generators of degree $n$,
are determined by the primitive generators on the factors of
$G/H$ of degrees up to $n$.

 First, the de Rham cohomology of $G/H$ can be identified with the
coset of the space of all closed $G$-invariant forms on $G/H$ by the
space of exterior derivatives of $G$-invariant forms on $G/H$. The
$G$-invariant forms on $G/H$ are given by Eq. (3.5) where the
coefficients $\omega _{A_1 \cdots A_n}$ are (1) constant, (2)
invariant under the adjoint action of $\H$, and (3) zero whenever any
of the indices $A_i$ corresponds to a generator of $\H$.

Second, we construct all closed forms obeying (1), including all
exact forms of the type (3.5), but ignoring properties (2) and (3)
temporarily. To do so, we exhibit a map (the transgression ${}^8$)
that lowers the degree but increases the rank.  Let $\Omega$ be a
closed
$G$-invariant form of rank 0 and degree $n$;  using the first
equation in (3.7), we see that $D(i_A\Omega)=0$. Applying the second
relation in (3.7), we further see that $i_A\Omega$ is exact as a form
on $G$
$$
i_A \Omega =  D \Omega _A
\qquad \qquad
\Omega _A =  DM _A  -{1 \over L^2} \Delta (i_A \Omega)
\eqno (3.8)
$$
The quadratic Casimir operator in Eq. (3.8) is invertible on
$i_A\Omega$ since it is invertible on the simple components of $\G$
and since the presence of invariant $U(1)$ components of $\G$ in $i_A
\Omega$ does not occur when $\Omega$ is a primitive form of degree
$n\geq 2$. From $\Omega _A$, we construct a new
\footnote{*}{Curly brackets denote symmetrization of the
corresponding indices.}
form $i_{\{A} \Omega _{B\}}$, which is again closed in view of
Eq. (3.7-8). This process may be continued into a hierarchy of
equations. For a {\it primitive generator} of degree $n$, the
hierarchy terminates at $m= [(n-1)/2]$ where forms of degree 0 are
encountered.
$$
\eqalign{
D(i_{B_1}\Omega ) =0
\quad \Rightarrow & \quad
i_{B_1} \Omega
= D\Omega _{B_1},
\cr
D(i_{\{B_{k+1}} \Omega _{B_1 \cdots B_k\}} )=0
\quad \Rightarrow & \quad
i_{\{B_{k+1}} \Omega _{B_1 \cdots B_k \}}
= D\Omega _{B_1 \cdots B_{k+1}},
\qquad \qquad
{\scriptstyle 1\leq k \leq m-1}
\cr
D(i_{\{B_{m+1}} \Omega _{B_1 \cdots B_m\}} )=0
\quad \Rightarrow & \quad
i_{\{B_{m+1}} \Omega _{B_1 \cdots B_m \}}
= D\Omega _{B_1 \cdots B_{m+1}} + d _{B_1 \cdots B_{m+1}}.
\cr}
\eqno (3.9)
$$
For odd $n=2m+1$, $d _{B_1 \cdots B_{m+1} }$ is a constant
$G$-invariant tensor, completely symmetric in its indices, and
$\Omega _{B_1 \cdots B_{m+1}}=0$.
 When $H=1$, we recover the cohomology of Lie groups of \S 3.1;
no extra conditions are needed in this case and  the invariant tensor
$d$ reproduces the generator of degree $2m+1$ on $G$, given
by Eq. (3.3).   When $n=2m+2$ is even,
$d _{B_1 \cdots B_{m+1} }=0$ and $\Omega _{B_1 \cdots B_{m+1}}$ is
constant and $\H$-invariant, but it is not necessarily
$\G$-invariant. The analysis of cohomology is thus reduced to the
analysis of constant tensors with certain invariance properties, a
problem that can be solved in terms of group characters.

Third, property (3) puts restrictions on the allowed
symmetric tensors : {\it the $d$-symbols must vanish whenever all of
its indices correspond to generators of $\H$}. This follows from a
repeated application of the map used in Eq. (3.8)
$$
d_{B_1\cdots B_m} = \prod _{i=1} ^{m-1} \biggl \{ - i_{\{B_i} {1
\over L^2} \Delta \biggr \}  \bigl ( i_{B_m\}} \Omega \bigr )
\eqno (3.10)
$$
Since the original $\Omega$ vanishes on $\H$, the $i_{B_i}$ to the
extreme left will always vanish when all indices $B_i$ correspond to
generators in $\H$. Thus, a non-zero $d$ tensor will correspond to a
primitive generator on $G/H$ only if it vanishes on $\H$.

Finally, one integrates the hierarchy of equations and obtains all
closed forms of the type (3.5)  obeying properties (1) and (2).
Property (3) is imposed explicitly and exact generators are discarded
by simple enumeration. The remaining forms are
precisely the de Rham cohomology generators of $G/H$.

\vglue1cm
\leftline{\bf 4. Special Effective Actions in Dimensions 1, 2, 3
and 4}
\vglue0.4cm

Cohomology of degree 1 arises when $G/H$ is not simply connected, and
the generators are easily found by duality with the homology cycles
on $G/H$. If $G/H$ has product factors of $U(1)$, there will be a
generator of degree 1 for each factor.

Primitive cohomology generators of degree 2 are completely determined
by the maximal number $r$ of Abelian $U(1)$ factors of $H$.
To each generator $T^{\alpha_l}$, there corresponds a primitive
generator
$$
\Omega ^{(l)} = -d(\theta ^{\alpha _l}) = - W_{\alpha _l}
\qquad \qquad
1\leq l\leq r
\eqno (4.1)
$$
where $W$ is the field strength associated with the composite
$\H$-valued gauge field $V$, defined in Eq. (2.3). These generators
span the first Chern class of $G/H$.

Primitive  generators of degree 3 are associated with
$G$-invariant symmetric rank 2 tensors $d_{BC}$ on $\G$ which vanish
on $\H$ and it is easy to integrate the formalism of \S
3.3. The primitive generators are Goldstone-Wilczek currents
${}^{16}$, gauged under the subgroup $H$ and given by
$$
\Omega = {1 \over 6} d_{ab} f_{bcd} \theta ^a \theta ^b \theta ^c +
{2\over 3}d_{a \beta} f_{\beta cd} \theta ^a \theta ^b \theta ^c
\eqno (4.2)
$$

A primitive generator $\Omega$ of degree 4 produces a hierarchy in
(3.9) where the $d$ tensor is absent, and where $\Omega _{B_1B_2}$
is a constant $\H$-invariant symmetric tensor on $\G$. As a form on
$G$, $\Omega$ must be exact, since there are no primitive generators
of even degree on a Lie group. Applying the procedures of \S 3.3, we
find that
$$
\Omega =  d Q \qquad \qquad Q= \half m_{\beta \lambda } f_{a \beta c}
\theta ^a \theta ^c \theta ^\lambda
+{ 1 \over 6} m_{\beta \lambda } f_{\alpha \beta \gamma} \theta ^\alpha
\theta ^\gamma \theta ^\lambda
\eqno (4.3)
$$
or $\Omega = m_{\alpha \beta} W_\alpha W_\beta $.
Here, $ W_\alpha = d\theta _\alpha +  f_{\alpha \gamma \delta}  \theta
_\gamma \theta _\delta/2$ are the components of the curvature form
$W$  and $m_{\alpha \beta}$ is any constant symmetric tensor,
invariant under the action of the adjoint representation  of $\H$.
These tensors are just the Cartan-Killing  forms on the simple
components of $\H$, and arbitrary coefficients on the $U(1)$
components of $\H$. The full cohomology of degree 4 (in the case where
$G/H$ is simply-connected) is given by
$$
\Omega = \sum _{k=1} ^q m_2 ^{(k)} W_\alpha ^{(k)} W_\alpha ^{(k)}
+ \sum _{l,m=1} ^r m_1 ^{(l,m)} W_{\alpha _l} W_{\alpha _m}
\eqno (4.4)
$$
The generators of $H^4 (G/H;\R)$ in the first sum belong to the
second Chern class evaluated on the
$\H$-valued  connection $V$ of (2.3) with components $\theta
^\alpha$, while the generators in the second sum arise from products
of generators belonging to the  first Chern class.  The form $Q$ in
(4.3) is a linear combination of Chern-Simons invariants
in 3 dimensions ${}^{9,10}$, evaluated on the composite gauge field
$V$ of (2.3).  The resulting invariant effective action
coincides with the Chern-Simons action evaluated on composite
connections.

Cohomology of degree 5 is the one relevant to actions in
4-dimensional space-time. According to the general procedure outlined
in \S 3.3, the primitive generators of degree 5 are associated with a
completely symmetric $G$-invariant tensor $d_{ABC}$ of rank 3 on
$\G$, which must vanish on $\H$. This tensor is precisely the one
encountered in the study of the chiral gauge anomaly ${}^{10}$ for
a gauge group $G$. Conversely, any non-zero $G$-invariant tensor
$d_{ABC}$, which vanishes on the subalgebra $\H$ produces a unique
primitive generator of $H^5(G/H;\R)$, given by
$$
\eqalign{
 \Omega = & {1 \over 480} \bigl \{
d_{a_1 bc} f_{ba_2a_3} f_{ca_4a_5}
+ 7 d_{a_1 b\gamma } f_{ba_2a_3} f_{\gamma a_4a_5} \cr
& ~~~~+ 16 d_{a_1 \beta \gamma } f_{\beta a_2a_3} f_{\gamma a_4a_5}
\bigr \}
\theta ^{a_1}\theta ^{a_2}\theta ^{a_3}\theta ^{a_4}\theta ^{a_5}
\cr}
\eqno (4.5)
$$
Notice that $d_{\alpha \beta \gamma}$ does not enter, and that for
$H=1$, we recover $\Omega _5$ of (3.3). All other generators of
$H^5(G/H;\R)$ are decomposable into linear combinations of products of
generators of degrees 1, 2, 3 and 4 which were already discussed
above. In particular, the coupling of the Goldstone-Wilczek current to
an Abelian composite gauge field strength $W_{\alpha _l}$ of (4.1) is
of this type and the corresponding action may be recast in four
dimensions as the Goldstone Wilczek current coupled to an Abelian
(composite) gauge field $\theta ^{\alpha _l}$.

An alternative procedure for obtaining the same primitive generators
is already familiar and was used for example in Ref. 1. In terms of
the
$\H$-valued gauge field $V$, the $\H$-covariant derivative  $D^{\H}$
of Eq. (2.3), and the trace representation for the $d$-symbol, we
obtain a form ${}^{12,13,17}$ $\Omega$ that indeed vanishes on $\H$,
given by
$$
 \Omega = \sum _j {1 \over 5!}~ d_3 ^j ~\tr _{\G_j} ~
\biggl [ \bigl ( U ^{-1} D^{\H} U \bigr ) ^5 - 5 W \bigl ( U ^{-1}
D^{\H} U \bigr )^3 +10 W ^2 \bigl (U ^{-1} D^{\H} U \bigr )\biggr ]
\eqno (4.6)
$$
Closure of this form is guaranteed by the fact that $d_{ABC}$
vanishes on $\H$. Also, this generator cannot be decomposed into a sum
of products of generators of lower degree that are well-defined on
$G/H$, and thus  $ \Omega$ is {\it primitive}.

\bigskip
\bigskip

\leftline{\bf 5. Quantization Conditions, fractionalized
statistics of solitons}

\bigskip

Different interpolations $\tilde \pi$ may be topologically
inequivalent; there is no natural way of choosing one
interpolation above another. The quantum action, however, is allowed
to change additively by integer multiples of $2\pi$ under change of
interpolating map${}^6$. The dependence of interpolation then
becomes invisible in the quantum theory provided the coupling
constants entering $\Omega $ are suitably quantized.

Precisely which coupling constants must be quantized depends upon
the topology of the space-time manifold $M_{n-1}$ and upon the
corresonding cohomology generator $\Omega$ in $H^n(G/H;\R)$. Two
interpolations may be viewed ${}^6$ as interpolations on
different balls $B_n^1$ and $B_n^2$ both with boundary $M_{n-1}$.
If the two interpolations are topologically inequivalent, then the
two balls cannot be continuously connected to each other, and the
difference $C_n = B_n^2 - B_n^1$ must form a non-contactible cycle
$C_n$ inside $G/H$.  As a general rule, quantization of the coupling
constant of $\Omega$ must occur when the integral of $\Omega$ on
the difference cycle $C_n$  is non-zero for at least one pair of
interpolations. Notice that in our formulation, quantization
conditions for WZW terms and Chern Simons terms appear on the same
footing.

If space-time is a sphere $S^{n-1}$, the difference cycle
is a sphere $C_n=S^n$, and {\it only the coupling constants for
primitive generators of $H^n(G/H;\R)$ must be quantized.} This
includes the cases of the WZW terms in two and
four dimensions, of the Chern-Simons action for {\it the semi-simple}
factors of $\H$ in three dimensions, and of the first Chern class in 1
dimension which is the original Dirac quantization condition for
magnetic monopoles.  Coupling constants for decomposable generators do
not have to be quantized however. This case includes in four
dimensions the coupling of the Goldstone Wilczek current to a
composite $U(1)$ field, or the Chern Simons action in three
dimensions associated with the Abelian factors in $\H$. If
space-time is not a sphere, then the quantization conditions are
different, and coupling constants for decomposable generators may
have to be quantized as well.

Further refinements of the quantization conditions may be required
depending on additional topological issues. We shall just consider an
example in which space-time is a sphere $S^4$, with $\pi _4(G)=0$ and
$\pi _4(H) \not=0$. For all simple groups $H$ we have $\pi _4(H)=0$,
except $\pi _4 (Sp(2p))={\bf Z}_2$.
Whenever $\pi_4(H)\not=0$, $H$ has a {\it discrete anomaly} [18],
and it can be shown that the
coupling constant of the corresponding WZW term of $H^5(G/H;{\bf
R})$ must be quantized in terms of {\it even} integers to
obtain a single-valued path integral ${}^1$.

The presence of the invariant actions discussed above in any
dimension of space-time has important physical consequences on the
dynamics of extended objects such as solitons. Witten showed in Ref.
6 that the integer coupling constant $N$ of the WZW term in four
dimensions can be viewed as the number of colors of an underlying
microscopic quark theory. Solitons in the Goldstone field then
acquire bosonic or fermionic statistics depending on whether $N$ is
even or odd${}^6$.

Wilczek and Zee showed in Ref. 11 that in three space-time dimensions
and $G/H = S^2$, solitons may acquire fractional spin and statistics
due to the presence of a Chern-Simons type term.  Our formalism
allows for a natural extension of their work to the case of any
symmetry breaking pattern $G/H$ when $H$ contains a certain number
$r$ of commuting $U(1)$ factors. Assuming $G$  simply connected,
there will be $r$ different types of solitons, each with its
independent conserved charge $g_l$, $l=1, \cdots , r$,
proportional to the space integral of the $U(1)$ field strength
$W_{\alpha _l}$ of (4.1). The corresponding effective action
(involving only the Chern Simons terms for the Abelian components of
$H$) may be read off directly from (4.4)
$$
S[\pi] = S_0[\pi]+ \sum _{l,m=1}^r m_1 ^{(l,m)} ~\int _{M_3} \theta
^{\alpha _l} d \theta ^{\alpha _m}
\eqno (5.1)
$$
Here $S_0[\pi]$ are manifestly $G$-invariant contributions to the
effective action which in particular would guarantee the existence of
solitons.  In view of the preceding discussion, the coupling
constants $m_1 ^{(l,m)}$ do not have to be quantized if the topology
of space-time is that of a 3-sphere. A soliton with charges $g_l =
\delta _{l,p}$ acquires a fractional spin proportional to
$m_1^{(p,p)}$ through the terms in (5.1). A soliton with charges $g_l
= \delta _{l,p} + \delta _{l,q}$ acquires a fractional spin,
which depends on $m_1^{(p,p)}$ and $m_1^{(q,q)}$, but also on a mixing
term $m_1^{(p,q)}=m_1^{(q,p)}$. A detailed analysis of this
fractionalization phenomenon is given in Ref. 3.

\bigskip
\bigskip

\leftline{\bf  Acknowledgements}

\bigskip

It is a pleasure to thank Steven Weinberg for collaboration on the
first part of this work. I have also benefited from helpful
conversations with Edddie Farhi, Chris Fronsdal, Jeff Rabin, Terry
Tomboulis and especially Steven Weinberg.

\bigskip
\bigskip

\leftline{\bf  References}

\bigskip

\item{1.} E. D'Hoker and S. Weinberg, Phys. Rev. {\bf D50} (1994) 605.

\item{2.} E. D'Hoker, ``Invariant Effective Actions, Cohomology of
Homogeneous Spaces and Anomalies'', UCLA/95/TEP/5 preprint (1995),
hep-th-95-02162, to appear in Nucl. Phys. B.

\item{3.} E. D'Hoker,  ``Soliton Spin Fractionalization in General
 2+1 Dimensional non-Linear Sigma Models'', UCLA/95/TEP/13 preprint
(1995); to appear.

\item{4.} S. Weinberg, Phys. Rev. {\bf 166} (1968) 1568;
		S. Coleman, J. Wess and B. Zumino, Phys. Rev. {\bf 177}
		(1969) 2239; C.G. Callan, S.~Coleman, J. Wess and B. Zumino,
		Phys. Rev. {\bf 177} (1969) 2247.

\item{5.} J. Wess and B. Zumino, Phys. Lett.  {\bf 37B} (1971) 95.

\item{6.} E. Witten,  Nucl. Phys. {\bf B223} (1983) 422, 433.

\item{7.} {\it Encyclopedic Dictionary of Mathematics},
                 S. Iyanaga and Y. Kawada, eds. (MIT Press, 1980).

\item{8.} H. Cartan, in {\it Colloque de Topologie, Centre Belge
de Recherches Math\'ematiques, Brussels 1950}, (G. Thone, 1950);
W. Greub, S. Halperin and R. Vanstone, {\it
	Connections, Curvature and Cohomology}, Vol III, (Acad. Press, 1976).

\item{9.} S. Chern, {\it Complex
Manifolds without Potential Theory}  (Springer Verlag, 1979).

\item{10.} see e.g. R. Jackiw, in {\it Current Algebra and Anomalies},
by S.B. Treiman, R. Jackiw, B. Zumino and E. Witten, Princeton Univ.
Press, 1985.

\item{11.} F. Wilczek and A. Zee, Phys. Rev. Lett. {\bf 51} (1983)
2250; F. Wilczek, {\it Fractional Statistics and Anyon
Superconductivity}, World Scientific, Singapore, 1990.

\item{12.} Y.-S. Wu, Phys. Lett. {\bf 153B} (1985) 70; B. De Wit,
C.M. Hull and M. Ro\^cek, Phys. Lett. 184B (1987) 233.
 C.M. Hull and B. Spence, Phys. Lett. {\bf 232 B} (1989) 204; I. Jack,
D.R. Jones, N. Mohammedi and H. Osborn, Nucl. Phys. {\bf B332} (1990)
359; C.M. Hull and B. Spence, Nucl. Phys. B353 (1991) 379.

\item{13.}  B. Zumino, in {\it Relativity, groups and Topology II:
Les Houches 1983}, B. De Witt, R. Stora, eds. (North Holland, 1984);
K.~Chou, H.Y.~Guo, K.~Wu and X.~Song, Phys. Lett. {\bf 134B} (1984)
67; J.~Manes, R.~Stora and B.~Zumino, Comm. Math. Phys. {\bf 102}
(1985) 157; J.~Manes, Nucl. Phys. {\bf B250} (1985) 369.

\item{14.} B.A. Dubrovin, A.T. Fomenko and S.P. Novikov, {\it
	Modern Geometry and Applications}, Vol III (Springer Verlag, 1990);
 M.~Spivak, {\it Differential Geometry, Vol. 5}, Publish or Perish, Inc.
Houston, 1975; S.I. Goldberg, {\it Curvature and Homology}, Dover
Publications, Inc., New York, 1982.

\item{15.}  S. Axelrod, Princeton Ph.D. thesis, unpublished (1991);
H.~Leutwyler, ``Foundations of Chiral Perturbation Theory",  Bern
preprint, BUTP-93/24, to appear in Annals of Physics;
 S.~Wu, J. Geom. Physics, {\bf 10} (1993) 381; J. M.~Figueroa-Farrill
and S.~Stanciu, {\it Equivariant Cohomology and Gauged Bosonic Sigma
Models}, QMW-PH-94-17, hep-th/9407149 preprint (1994).

\item{16.} J. Goldstone and F. Wilczek, Phys. Rev. Lett. {\bf 47}
(1981) 986; E. D'Hoker and J. Goldstone, Phys. Lett. {\bf
158B} (1985) 429.

\item{17.} E. D'Hoker and E. Farhi, Nucl. Phys. {\bf B248} (1984)
59, 77.

\item{18.} E. Witten, Phys. Lett. {\bf 117B} (1982) 324.

\end